# ЭФФЕКТЫ ФОКУСИРОВКИ АТОМОВ, ЭМИТИРОВАННЫХ С ГРАНИ (001) Ni, С РАЗРЕШЕНИЕМ ПО УГЛАМ И ЭНЕРГИИ

© 2017 г.   В. Н. Самойлов, А. И. Мусин

*Физический факультет, Федеральное государственное бюджетное образовательное учреждение высшего образования "Московский государственный университет имени М.В. Ломоносова", Москва, Россия. E-mail: samoilov@polly.phys.msu.ru*

В работе с помощью компьютерного моделирования методом молекулярной динамики исследованы особенности фокусировки и перефокусировки атомов, распыленных с поверхности грани (001) Ni. Показано, что при переходе кристалла из парамагнитного в ферромагнитное состояние наблюдается уменьшение числа фокусированных и перефокусированных распыленных атомов. Изучена эволюция распределений распыленных атомов с одновременным разрешением по энергии и полярному углу при изменении азимутального угла и энергии связи атомов на поверхности.

## ВВЕДЕНИЕ

К настоящему времени опубликовано большое число исследований анизотропии двумерного углового распределения атомов, распыленных с поверхности низкоиндексных граней монокристалла под действием ионной бомбардировки, являющейся одним из сложных эффектов, отражающих анизотропию структуры поверхности кристаллов. Картина углового распределения распыленных атомов чувствительна к типу облучаемой ионами грани кристалла [1–3]. Очень важным в теории распыления является вопрос, какой механизм является ответственным за формирование особенностей двумерного углового распределения атомов, распыленных с поверхности монокристалла, с разрешением по энергии. В [4–7] представлено обсуждение механизмов фокусировки атомов, распыленных с поверхности низкоиндексных граней монокристалла, по полярному и азимутальному углам, в том числе с разрешением по энергии.

В ряде работ была предложена своя классификация механизмов фокусировки распыленных атомов. В частности, в [8] представлена классификация цепочек столк-новений, приведших к распылению атома, с разделением на F (фокусированные), C (дефокусированные), CF и FC (смешанные директоны), введенные в [8]. В [9] предложена классификация распыленных атомов, формирующих анизотропию двумерного углового распределения, по разнице n номеров атомных слоев кристалла для атома, от которого получил импульс распыленный атом, и первоначального положения распыленного атома, $\Delta_0$, $\Delta_1$, $\Delta_2$, $\Delta_3$. Для описания формирования особенностей полярного углового распределения распыленных атомов с разрешением по энергии в [5, 10] предложена классификация распыленных атомов с разделением на сильно блокированные по полярному углу и остальные распыленные атомы. Для сильно блокированных распыленных атомов отклонение по полярному углу в направлении нормали к поверхности при эмиссии атома вследствие его рассеяния на атомах поверхности больше отклонения в сторону от нормали к поверхности вследствие преломления на плоском потенциальном барьере.

Отметим также ряд важных для настоящей работы результатов исследований. В [11] было обнаружено, что при малых энергиях бомбардирующих ионов распределение распыленных атомов по полярному углу в каскаде столкновений, пересекающем поверхность, не имеет максимумов вблизи плотноупакованных направлений <011>. При этом в угловом распределении распыленных атомов наблюдались максимумы эмиссии, которые по своей угловой ширине и направлениям формирования соответствовали наблюдаемым пятнам Венера. В [12] было показано, что блокировка траекторий эмитируемых атомов в сторону меньших полярных углов и по азимутальному углу является одним из основных механизмов формирования наблюдаемых пятен Венера.

В расчетах эмиссии атомов с поверхности граней (001) Ni и (111) Ni, в частности, с разрешением по энергии, наблюдались максимумы в двумерном угловом распределении распыленных атомов, которые по своей угловой ширине и направлениям формирования соответствовали экспериментально наблюдаемым максимумам эмиссии – пятнам Венера [5]. Таким образом, формирование экспериментально наблюдаемых пятен Венера в двумерном угловом распределении атомов, распыленных с поверхности монокристалла, оказалось возможным объяснить действием только поверхностного механизма фокусировки. На стадии эмиссии происходит сильное перераспределение вылетающих атомов по углам и энергии, такое, что, стадия эмиссии играет важную роль в формировании углового и энергетического распределений распыленных атомов.

Для описания формирования особенностей распределения распыленных атомов по полярному и азимутальному углам с разрешением по энергии в [6, 7] предложена классификация распыленных атомов с разделением на "собственные" по азимутальному углу и "несобственные" атомы: фокусированные и перефокусированные. Для несимметричных относительно направления <100> интервалов азимутального угла φ формирование сигнала распыленных атомов происходит за счет "собственных" атомов, начальный угол вылета которых $φ_0$ принадлежит интервалу углов φ, и фокусировки "несобственных" атомов: фокусированных атомов, рассеянных на ближайшем атоме линзы из двух ближайших к эмитируемому атому атомов поверхности, и перефокусированных атомов, рассеянных на дальнем атоме линзы (рис. 1). Для фокусированных атомов угол φ и угол $φ_0$ лежат по одну сторону от направления <010> на центр линзы из двух ближайших к эмитируемому атому атомов поверхности, для перефокусированных атомов – по разные стороны от этого направления. Таким образом, фокусировка атомов идет к центру линзы из двух атомов, а перефокусировка – через центр линзы из двух атомов поверхности. Эффект перефокусировки был обнаружен в [4, 11] и исследован в ряде работ, например, в [6, 7].

В настоящей работе ставилась задача изучить вклад фокусированных и перефокусированных атомов в формирование распределений распыленных атомов по углам и энергии, изучить вопрос о выделении перефокусированных атомов в общем сигнале эмитированных атомов и исследовать изменения распределения распыленных атомов с разрешением по углам и энергии при переходе грани (001) Ni из парамагнитного (p) состояния в ферромагнитное (f).

## МОДЕЛЬ РАСЧЕТА

Расчеты были проведены для эмиссии атомов с поверхности грани (001) Ni. Поверхность кристалла моделировалась 20 атомами поверхности, ближайшими к узлу решетки, из которого происходила эмиссия атома (модель 21 атома). Фрагмент грани (001) Ni, используемый в расчете, представлен на рис. 1. Подобная модель использовалась в наших работах [7, 13].

Для расчета эмиссии атомов использовался метод молекулярной динамики.

Взаимодействие эмитируемого атома с атомами поверхности в модели описывалось потенциалом отталкивания, а на достаточно большом удалении атома от поверхности был введен плоский потенциальный барьер. В качестве потенциала взаимодействия атом–атом был использован потенциал Борна–Майера:

$$U(r) = A\exp(-r/b) \tag{1}$$

с параметрами $A = 23853.96$ эВ и $b = 0.196$ Å для взаимодействия двух атомов Ni из работы [14]. Энергия связи составляла 4.435 эВ.

Атом выбивался из узла на поверхности с энергией $E_0$ под углами $\vartheta_0$ (начальный полярный угол, отсчитывался от нормали к поверхности) и $\varphi_0$ (начальный азимутальный угол, $\varphi_0 = 90°$ соответствовал направлению <010> на центр линзы из двух ближайших к эмитируемому атому атомов поверхности). Начальная энергия $E_0$ изменялась от 0.5 эВ до 100 эВ. Шаг по $E_0$ составлял 0.01 эВ. Шаг по $\varphi_0$ был равен 0.5°, шаг по $1 - \cos\vartheta_0$ составлял 1/450. Было использовано начальное распределение эмитируемых атомов по углам и энергии $\cos\vartheta_0/E_0^2$ [15, 16]. Таким образом, распределение эмитируемых атомов по начальному азимутальному углу $\varphi_0$ было изотропным.

Считалось, что распыление происходит только за счет атомов поверхностного слоя. Это допущение вполне оправдано ввиду того, что для мишеней, состоящих из средних по массе и тяжелых атомов, вклад атомов поверхностного слоя в распыление является доминирующим (88,6 % для случая ионной бомбардировки Cu [17], 82% для случая ионной бомбардировки Mo [18]). Обсуждение некоторых особенностей и корректности модели, используемой в настоящей работе, приведено также в работе [4].

Расчеты были выполнены с использованием ресурсов суперкомпьютерного комплекса МГУ "Ломоносов" [19, 20].

## РЕЗУЛЬТАТЫ И ИХ ОБСУЖДЕНИЕ

### *Об изменении числа фокусированных и перефокусированных распыленных атомов при магнитном фазовом переходе*

В настоящей работе проведено моделирование эмиссии атомов с грани (001) Ni при переходе из парамагнитного (p) состояния в ферромагнитное (f). Фазовый переход моделировался изменением потенциала взаимодействия атом–атом и изменени-ем поверхностной энергии связи (смотри также [21–24]).

В ферромагнитном состоянии взаимодействие атомов включает в себя спиновое обменное взаимодействие. Для атомов Ni оно было рассчитано в работе [25]. Добавка к потенциалу взаимодействия двух атомов Ni в ферромагнитном состоянии отрицательна:

$$\Delta U_f(r) = -5.16 \exp(-0.8112 r^2). \qquad (2)$$

Здесь $\Delta U_f(r)$ измеряется в эВ, а $r$ – в Å. Кроме того при переходе в f-состояние увеличивается энергия связи атомов $E_b$. Увеличение энергии связи атома при p–f переходе для грани (001) Ni было рассчитано в работе [21] и оценивается в 5%. Таким образом, в f-состоянии потенциал отталкивания двух атомов Ni менее жесткий, а энергии связи атома на поверхности больше, чем в p-состоянии.

При p–f переходе наблюдается уменьшение числа всех распыленных, фокусированных и перефокусированных по азимутальному углу распыленных атомов на 4.49% (5.48%), 9.04% (6.07%) и 8.44% (8.91%) соответственно при изменении потенциала взаимодействия атом–атом (изменении энергии связи). Таким образом, сигналы фокусированных и перефокусированных атомов оказываются более чувствительными, чем сигнал всех распыленных атомов, к изменению магнитного состояния мишени. Под влиянием каждого из двух факторов вклад фокусированных и перефокусированных атомов в распыление уменьшается на 2.62% (0.35%) и 0.28% (0.25%) соответственно. Таким образом, для этих групп атомов изменения потенциала взаимодействия и энергии связи дают вклад в одну сторону (в сторону уменьшения). Это можно объяснить тем, что при p–f переходе потенциал взаимодействия атом–атом становится менее жестким, а энергия связи увеличивается, что затрудняет вылет этих групп атомов. Напротив, вклад "собственных" по азимутальному углу атомов увеличивается на 2.90% (0.60%) при p–f переходе.

### *Об эволюции распределений фокусированных и перефокусированных распыленных атомов с изменением угла наблюдения*

В [4, 5, 10, 26] был обнаружен и исследован немонотонный сдвиг максимума полярного углового распределения распыленных атомов с ростом их энергии. В основе эффекта лежит конкуренция двух факторов: блокировки эмитируемых атомов в сторону нормали к поверхности в процессе вылета и преломления на плоском потенциальном барьере. Такой сдвиг максимума полярного углового распределения распыленных атомов с ростом их энергии также наблюдался экспериментально [27].

В распределениях с одновременным разрешением по энергии и полярному углу для фиксированных интервалов углов φ отчетливо различаются отдельные хребты – максимумы распределений для фокусированных и перефокусированных атомов (рис. 2). Верхний хребет образован в основном фокусированными атомами, нижний – только перефокусированными атомами. Максимум распределения перефокусированных атомов наблюдается в области энергии и полярных углов, при которых нет вылета других групп атомов. Таким образом, в экспериментах с разрешением по углам и энергии оказывается принципиально возможным выделить отдельно сигнал только перефокусированных атомов.

Показано, что отчетливо различающиеся отдельные хребты – максимумы распределений для фокусированных и перефокусированных атомов – смещаются в сторону меньших полярных углов при сдвиге азимутального угла от центра линзы. Обнаружено, что при таком изменении азимутального угла происходит уменьшение максимальной энергии перефокусированных атомов и фокусированных атомов, формирующих низкоэнергетическую часть распределения. При этом наблюдается расширение области углов $φ_0$ вблизи направления на центр линзы, при которых не происходит распыления фокусированных и перефокусированных атомов, и соответственно сужение областей углов $φ_0$, при которых происходит распыление фокусированных и перефокусированных атомов.

При p–f переходе обнаружены заметные сдвиги максимумов распределений фокусированных и перефокусированных атомов в сторону больших энергий и больших полярных углов (рис. 2). Наблюдается также значительное уменьшение области тени для Ni в ферромагнитном состоянии.

### *Условия лучшего разрешения максимумов распыленных фокусированных и перефокусированных атомов*

Обнаружено, что при увеличении значения энергии связи максимумы распределений фокусированных и перефокусированных атомов также заметно смещаются в сторону больших полярных углов и больших значений энергии. Максимум перефоку-сированных атомов смещается более чем на 1.2 эВ при увеличении энергии связи менее чем на 0.6 эВ. Обсуждается вопрос о лучшем разрешении максимумов фокусированных и перефокусированных атомов. Для фиксированных интервалов полярного угла эти максимумы может разделять 2-4 эВ и более при ширине максимумов менее 1-2 эВ (рис. 3), а для фиксированных интервалов энергии они могут отстоять на 15-40° при достаточно малой угловой ширине. Анализ результатов показывает, что, по-видимому, нужно выбирать материалы с большей энергией связи для лучшего разрешения максимумов распыленных фокусированных и перефокусированных атомов.

## ВЫВОДЫ

С помощью модели молекулярной динамики исследованы особенности фокусировки и перефокусировки атомов, эмитированных с поверхности грани (001) Ni, по азимутальному углу при формировании интегрального числа распыленных атомов и распределений распыленных атомов с разрешением одновременно по полярному углу и энергии.

Показано, что при p–f переходе наблюдается уменьшение числа всех распыленных, фокусированных и перефокусированных по азимутальному углу распыленных атомов. При этом сигналы фокусированных и перефокусированных атомов оказываются более чувствительными, чем сигнал всех распыленных атомов, к изменению магнитного состояния мишени.

Обнаружено, что при сдвиге азимутального угла наблюдения от центра линзы происходит уменьшение максимальной энергии перефокусированных атомов и фокусированных атомов, формирующих низкоэнергетическую часть распределения. При этом наблюдается расширение области углов $\varphi_0$ вблизи направления на центр линзы, при которых не происходит распыления фокусированных и перефокусированных атомов.

Выявлено, что максимумы фокусированных и перефокусированных атомов оказались чувствительны к изменению энергии связи и магнитного состояния мишени. С увеличением энергии связи происходит сильное смещение максимумов фокусированных и перефокусированных атомов в сторону больших энергий.

Обсуждается вопрос о лучшем разрешении максимумов фокусированных и перефокусированных атомов. Анализ результатов показал, что, по-видимому, нужно выбирать материалы с большей энергией связи для лучшего разрешения максимумов распыленных фокусированных и перефокусированных атомов.

## СПИСОК ЛИТЕРАТУРЫ

# FOCUSING EFFECTS FOR ATOMS SPUTTERED FROM THE (001) Ni FACE WITH ENERGY AND ANGULAR RESOLUTION

**V. N. Samoilov, A. I. Musin**

*Faculty of Physics, M. V. Lomonosov Moscow State University, Moscow, 119991 Russia*

*e-mail: samoilov@polly.phys.msu.ru*

Peculiarities of the focusing and overfocusing of atoms sputtered from the surface of (001) Ni face are studied with the use of molecular dynamics computer simulations. It is discovered that the numbers of focused and overfocused sputtered atoms decrease under the transition of the (001) Ni face from the paramagnetic to the ferromagnetic state. The evolution of the two-dimensional energy and polar angle resolved distributions of sputtered atoms when changing the azimuthal angle and the binding energy of atoms at the surface is investigated.

ПОДПИСИ К РИСУНКАМ

**Рис. 1.** Рассеяние атома при вылете с поверхности и классификация эмитированных атомов по азимутальному углу φ (*а*). Представлена только линза, состоящая из двух атомов: ближайших к вылетающему атому соседей в плоскости поверхности. Фрагмент грани (001) Ni, используемый в расчете (модель 21 атома) (*б*). Вылет атома происходит из узла 1. Показана линза из двух атомов 2 и 3 – ближайших к вылетающему атому соседей в плоскости поверхности, и атом 4, на которых происходит рассеяние эмитируемого атома.

**Рис. 2.** Распределения распыленных атомов одновременно по 1 – cosϑ и энергии *E* для интервала азимутальных углов φ [76.5$^o$, 79.5$^o$] при эмиссии с грани (001) Ni в парамагнитном состоянии (*а*) и ферромагнитном состоянии (*б*). Верхний хребет образован в основном фокусированными атомами, нижний – только перефокусированными атомами.

**Рис. 3.** Распределения распыленных атомов по энергии *E* при эмиссии с грани (001) Ni для полярных углов вылета ϑ [49.9$^o$, 51.5$^o$] и интервала азимутальных углов φ [76.5$^o$, 79.5$^o$]. Значение энергии связи 4.435 эВ (*а*) и 5 эВ (*б*). Левый максимум образован в основном фокусированными атомами, правый – только перефокусированными атомами.

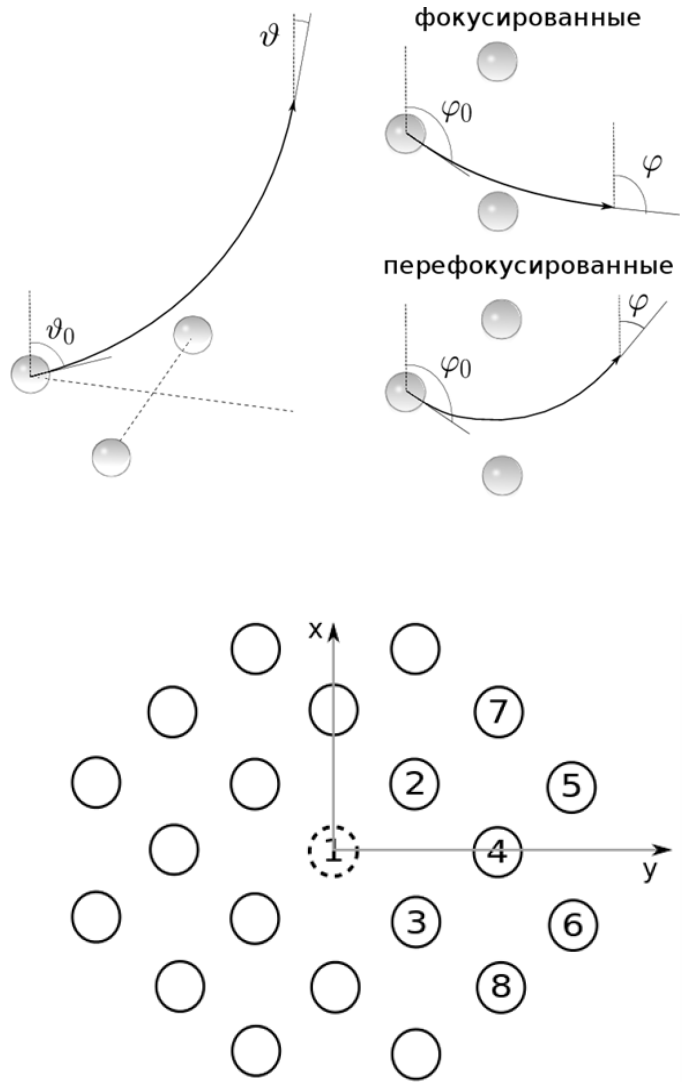

**Рис. 1.** Рассеяние атома при вылете с поверхности и классификация эмитированных атомов по азимутальному углу φ (*а*). Представлена только линза, состоящая из двух атомов: ближайших к вылетающему атому соседей в плоскости поверхности. Фрагмент грани (001) Ni, используемый в расчете (модель 21 атома) (*б*). Вылет атома происходит из узла 1. Показана линза из двух атомов 2 и 3 – ближайших к вылетающему атому соседей в плоскости поверхности, и атом 4, на которых происходит рассеяние эмитируемого атома.

В. Н. Самойлов, А. И. Мусин
ЭФФЕКТЫ ФОКУСИРОВКИ АТОМОВ, ЭМИТИРОВАННЫХ С ГРАНИ (001) Ni,
С РАЗРЕШЕНИЕМ ПО УГЛАМ И ЭНЕРГИИ

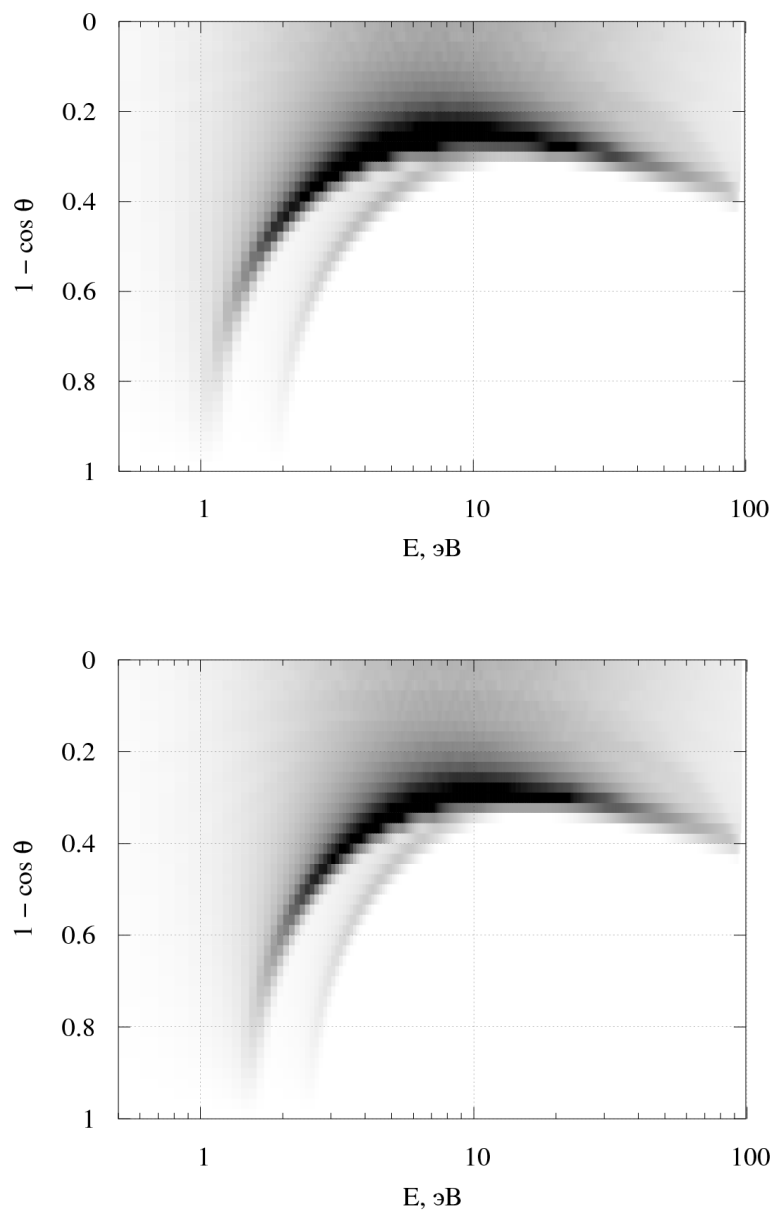

**Рис. 2.** Распределения распыленных атомов одновременно по 1 − cosϑ и энергии $E$ для интервала азимутальных углов φ [76.5°, 79.5°] при эмиссии с грани (001) Ni в парамагнитном состоянии (*а*) и ферромагнитном состоянии (*б*). Верхний хребет образован в основном фокусированными атомами, нижний – только перефокусированными атомами.

В. Н. Самойлов, А. И. Мусин

ЭФФЕКТЫ ФОКУСИРОВКИ АТОМОВ, ЭМИТИРОВАННЫХ С ГРАНИ (001) Ni,
С РАЗРЕШЕНИЕМ ПО УГЛАМ И ЭНЕРГИИ

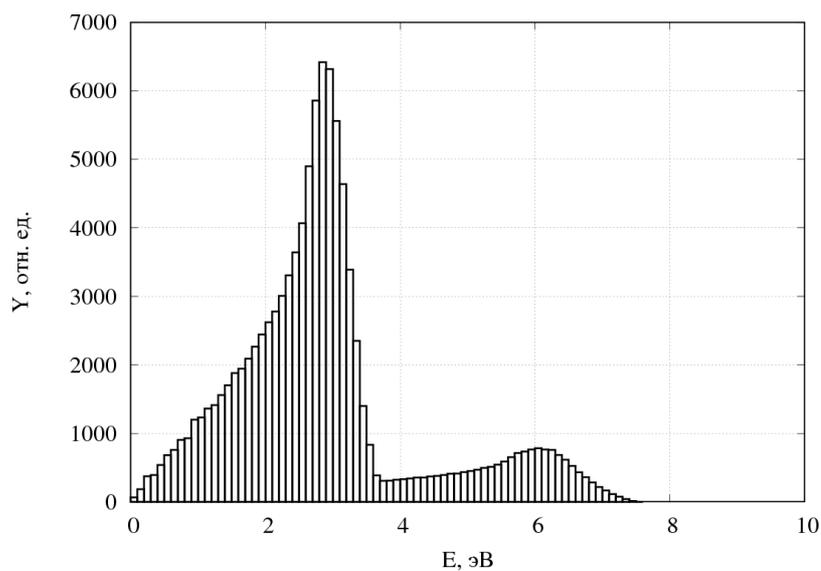

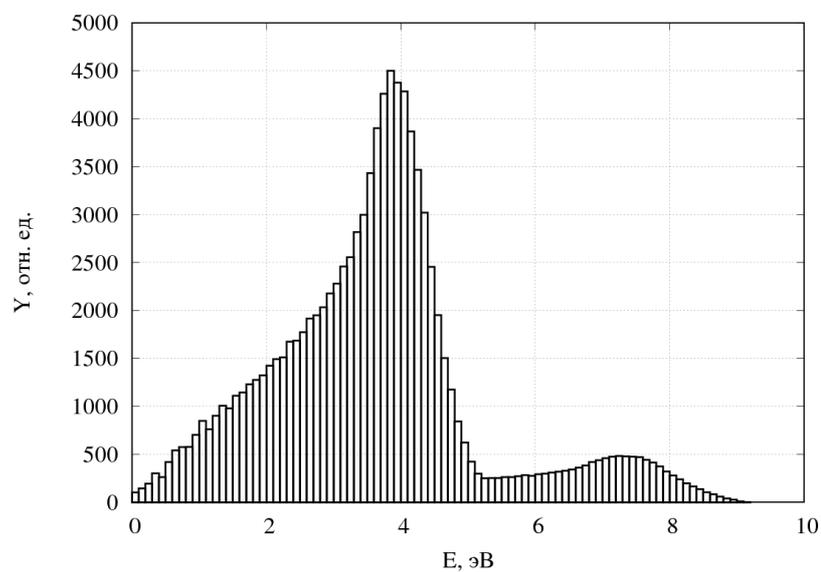

**Рис. 3.** Распределения распыленных атомов по энергии *E* при эмиссии с грани (001) Ni для полярных углов вылета ϑ [49.9°, 51.5°] и интервала азимутальных углов φ [76.5°, 79.5°]. Значение энергии связи 4.435 эВ (*а*) и 5 эВ (*б*). Левый максимум образован в основном фокусированными атомами, правый – только перефокусированными атомами.

В. Н. Самойлов, А. И. Мусин
ЭФФЕКТЫ ФОКУСИРОВКИ АТОМОВ, ЭМИТИРОВАННЫХ С ГРАНИ (001) Ni,
С РАЗРЕШЕНИЕМ ПО УГЛАМ И ЭНЕРГИИ